\documentstyle[12pt,epsfig,axodraw]{article}
\usepackage{epsfig}
\newcommand{\mysection}{\setcounter{equation}{0}\section}

\def\beq{\begin{equation}}
\def\eeq{\end{equation}}
\def\beqa{\begin{eqnarray}}
\def\eeqa{\end{eqnarray}}

\newlength{\dinwidth} \newlength{\dinmargin}
\begin{document}
\begin{center}
{\Large \bf Two-loop QCD calculations in the eikonal approximation}
\end{center}
\vspace{2mm}
\begin{center}
{\large Nikolaos Kidonakis\footnote{Presented at DPF2002, Williamsburg, 
Virginia, May 24-28, 2002}}\\
\vspace{2mm}
{\it Department of Physics\\
Southern Methodist University\\
Dallas, TX 75275-0175} \\
\vspace{2mm}
and \\
\vspace{2mm}
{\it Department of Physics and Astronomy\\
University of Rochester\\
Rochester, NY 14627-0171} \\
\end{center}

\begin{abstract}
I discuss recent progress in the calculation of
two-loop QCD corrections in the eikonal approximation.
I present specific results for the UV structure
of the corrections.
\end{abstract}

\thispagestyle{empty} \newpage \setcounter{page}{2}

\mysection{Introduction}

The eikonal approximation has been extremely useful
in calculations of QCD hard scattering cross sections,
including both fixed-order and resummed calculations.
It is valid for the emission and absorption of soft gluons from partons
in the hard scattering. When the gluon momentum goes to zero
the usual Feynman rules simplify as follows (Figure 1):

\beq
{\bar u}(p) (-i \gamma^{\mu}) \frac{i (p\!\!/-k\!\!/)}{(p-k)^2
+i\epsilon} \rightarrow {\bar u}(p) \gamma^{\mu} 
\frac{p\!\!/}{-2p\cdot k+i\epsilon}
={\bar u}(p)  
\frac{v^{\mu}}{- v\cdot k+i\epsilon}
\eeq
with $v$ a dimensionless vector, $p \propto v$. 

\vspace*{-1cm}
\begin{figure}[htb]
\begin{center}
\begin{picture}(120,120)(0,0)
\Line(0,80)(0,90)
\Line(3,80)(3,90)
\ArrowLine(0,85)(50,85)
\ArrowLine(50,85)(100,85)
\Vertex(50,85){2}  
\Gluon(50,25)(50,85){2}{8}
\Text(20,75)[c]{$p-k$}
\Text(80,75)[c]{$p$}
\LongArrow(65,35)(65,50)
\Text(85,40)[c]{$ k \rightarrow 0$}
\end{picture}
\end{center}
\vspace*{-1cm}
\caption{\label{eikonal}  Eikonal approximation}
\end{figure}
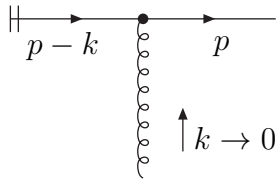

The eikonal approximation is used to derive the soft part
of the cross section in standard fixed-order QCD calculations \cite{BHJO}, 
and it has been used to calculate the high energy behavior of elastic
quark-quark scattering, (near) forward scattering amplitudes, 
and wide-angle elastic scattering \cite{BotSt,SoSt,Korchem,KaKtStW}. 
In addition, the eikonal approximation is a useful tool in developing
resummations for a variety of partonic processes.

Here we are mainly interested in the ultraviolet (UV) pole structure
of loop diagrams involving eikonal lines, because the UV poles play 
a direct role in renormalization group evolution equations that are used in 
threshold resummations \cite{KS,KOS,LOS,LSV}. 
At present the UV results are known to one loop and have been used
in state-of-the-art NLL resummations and derivative NNLO-NNLL
calculations for many processes [6-12]. 
To go beyond this level of accuracy two loop eikonal 
calculations are needed. Of course the study of the two-loop
UV structure of eikonal diagrams has intrinsic theoretical interest as well.
We note that there has also been recent progress 
in ordinary (i.e. not eikonal) two loop QCD calculations 
\cite{WGQCD,QCDSM,BFD} that will be 
needed in future fixed-order NNLO calculations.

We will limit our discussion here to eikonal lines 
that represent massive partons.
This is useful for heavy quark production, for example.
Studies for massless partons will be presented in future work.

\mysection{UV poles at one loop for eikonal diagrams with 
\newline massive partons}
 
We begin by reviewing one-loop results in the eikonal approximation
using the axial gauge. A representative one-loop vertex correction
diagram is given in Figure 2.

The general axial gauge gluon propagator is given by
\begin{equation}
D^{\mu \nu}(k)=\frac{-i}{k^2+i\epsilon} N^{\mu \nu}(k), \quad
N^{\mu \nu}(k)=g^{\mu \nu}-\frac{n^{\mu}k^{\nu}+k^{\mu}n^{\nu}}{n \cdot k}
+n^2\frac{k^{\mu}k^{\nu}}{(n \cdot k)^2},
\end{equation}
with $n^{\mu}$ the axial gauge-fixing vector.

The propagator for a quark, antiquark, or gluon eikonal line is
$i/(\delta v \cdot k +i \epsilon)$ with $\delta=+1 (-1)$ when the 
momentum $k$ flows in the same (opposite) direction as the dimensionless
vector $v$. The interaction gluon-eikonal line vertex for a quark or 
antiquark eikonal line
is $-i g_s (T_F^c)_{ba} v^{\mu} \Delta$ with $\Delta=+1(-1)$ for a quark
(antiquark), $g_s^2=4 \pi \alpha_s$,
and $T_F^c$ the generators of SU(3) in the fundamental 
representation. 
The interaction gluon-eikonal line vertex for a gluon eikonal line is
$-i g_s f^{abc} v^{\mu} \Delta$ with $\Delta=-i(+i)$ for a gluon located 
below (above) the eikonal line, $f^{abc}$ the totally antisymmetric
SU(3) invariant tensor, and we read the color indices $a,b,c$ anticlockwise.

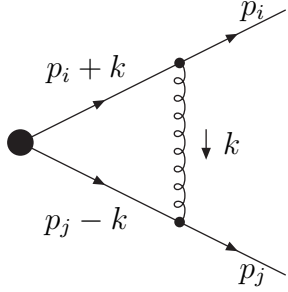
\begin{figure}[htb]
\begin{center}
\begin{picture}(120,120)(0,0)
\Vertex(0,50){5}
\ArrowLine(0,50)(60,80)
\ArrowLine(60,80)(100,100)
\Vertex(60,80){2}  
\Gluon(60,80)(60,20){2}{8}
\Text(25,78)[c]{$p_i+k$}
\Text(88,100)[c]{$p_i$}
\LongArrow(70,55)(70,45)
\Text(80,50)[c]{$k$}
\ArrowLine(0,50)(60,20)
\ArrowLine(60,20)(100,0)
\Vertex(60,20){2}
\Text(25,20)[c]{$p_j-k$}
\Text(88,0)[c]{$p_j$}
\end{picture}
\end{center}
\caption{\label{oneloop}  One-loop diagram}
\end{figure}

We denote the kinematic (color independent) 
part of the $n$-loop correction to the vertex, 
with the virtual gluon linking lines 
$v_{i}$ and $v_{j}$, as $\omega^{(n)}_{ij}(v_{i},v_{j})$.
The one-loop expression for $\omega_{ij}$ is then
\beqa   
\omega_{ij}^{(1)}(v_{i},v_{j})&\equiv&
{g}_{s}^2\int\frac{d^n k}{(2\pi)^n}\frac{-i}{k^2+i\epsilon}
N^{\mu \nu}(k) \frac{\Delta_{i} \: v_{i}^{\mu}}
{\delta_{i} v_{i} \cdot k+i\epsilon}\;
\frac{\Delta_{j} \:v_{j}^{\nu}} {\delta_{j}v_{j} \cdot k+i\epsilon}
\nonumber\\ &=&
{g}_{s}^2\int\frac{d^n k}{(2\pi)^n}\frac{-i}{k^2+i\epsilon}
\left\{\frac{\Delta_{i} \: \Delta_{j} \:v_{i}{\cdot}v_{j}}{(\delta_{i}v_{i}
{\cdot}k+i\epsilon)
(\delta_{j}v_{j}{\cdot}k+i\epsilon)}\right. 
\nonumber\\ &&
\left.{}-\frac{\Delta_{i} \, v_{i}{\cdot}n}{(\delta_{i}v_{i}
{\cdot}k+i\epsilon)}
\frac{P}{(n{\cdot}k)}-\frac{\Delta_{j} \, v_{j}{\cdot}n}{(\delta_{j}v_{j}
{\cdot}k+i\epsilon)}
\frac{P}{(n{\cdot}k)}+n^2\frac{P}{(n{\cdot}k)^2}\right\},         
\label{omega1}
\eeqa
where $P$ stands for principal value.
We may rewrite (\ref{omega1}) as
\beq
\omega^{(1)}_{ij}(v_{i},v_{j})= 
{\cal S}_{ij}^{(1)}
\left[I_1^{(1)}(v_i, v_j)
+I_2^{(1)}(v_i, v_j, n)
+I_3^{(1)}(n^2)\right] \, ,
\label{omegaI}
\eeq
where
\beqa
I_1^{(1)}&=&{g}_{s}^2\int\frac{d^n k}{(2\pi)^n} \frac{-i}{k^2+i\epsilon}\,
\frac{v_{i}{\cdot}v_{j}}{(v_{i}{\cdot}k+i\epsilon)(v_{j}{\cdot}k+i\epsilon)}
\nonumber \\
I_2^{(1)}&=&{g}_{s}^2\int\frac{d^n k}{(2\pi)^n}\frac{i}{k^2+i\epsilon}
\left(\frac{v_{i}{\cdot}n}{v_{i}{\cdot}k+i\epsilon}
+\frac{v_{j}{\cdot}n}{v_{j}{\cdot}k+i\epsilon}\right)\frac{P}{(n{\cdot}k)}
\nonumber \\
I_3^{(1)}&=&{g}_{s}^2\int\frac{d^n k}{(2\pi)^n}\frac{-i}{k^2+i\epsilon}\,
n^2\frac{P}{(n{\cdot}k)^2} 
\eeqa
and where ${\cal S}_{ij}^{(1)}$ is an  overall sign
\beq
{\cal S}_{ij}^{(1)}=\Delta_i \: \Delta_j \: \delta_i \: \delta_j.
\eeq
For the integrals when both $v_i$ and $v_j$ refer to massive quarks we have
(with $\varepsilon=4-n$) \cite{KS}
\beqa
I_1^{(1)\,\rm{UV \;pole}}
&=&\frac{\alpha_s}{\pi}\frac{1}{\varepsilon} L_{\beta} \, ,
\nonumber\\
I_2^{(1)\,\rm{UV \;pole}}&=&\frac{\alpha_s}{\pi}\frac{1}{\varepsilon} 
(L_i+L_j) \, ,
\nonumber\\
I_3^{(1)\,\rm{UV \;pole}}&=&-\frac{\alpha_s}{\pi}\frac{1}{\varepsilon} \, ,
\eeqa
where the $L_\beta$ is the  velocity-dependent
eikonal function
\begin{equation}
L_{\beta}=\frac{1-2m^2/s}{\beta}\left(\ln\frac{1-\beta}{1+\beta}
+\pi i \right)\, ,
\label{Lb}
\end{equation}
with $\beta=\sqrt{1-4m^2/s}$.
The $L_i$ and $L_j$ are rather complicated functions of
the gauge vector $n$. Their
contributions are cancelled by the inclusion of self energies.
Then, using Eq. (\ref{omegaI}), we get
\beq       
\omega_{ij}^{(1)\; \rm{UV \;pole}}(v_{i},v_{j})=
{\cal S}_{ij}^{(1)} \, \frac{\alpha_{s}}{\pi\varepsilon}
\left[L_{\beta} + L_i + L_j -1 \right].
\label{omegaheavy}
\eeq

\mysection{Leading UV poles at two loops for eikonal 
diagrams with massive partons}

We now continue with a calculation of the two-loop diagram
in Figure 3.

\begin{figure}[htb]
\begin{center}
\begin{picture}(120,120)(0,0)
\Vertex(0,60){5}
\ArrowLine(0,60)(60,85)
\ArrowLine(60,85)(100,102)
\ArrowLine(100,102)(120,110)
\Vertex(60,85){2}
\Vertex(60,35){2}  
\Gluon(60,85)(60,35){2}{8}
\Text(15,90)[c]{$p_i+k_1+k_2$}
\Text(75,105)[c]{$p_i+k_1$}
\Text(110,115)[c]{$p_i$}
\LongArrow(70,65)(70,55)
\Text(80,60)[c]{$k_2$}
\ArrowLine(0,60)(60,35)
\ArrowLine(60,35)(100,18)
\ArrowLine(100,18)(120,10)
\Vertex(100,102){2}
\Vertex(100,18){2}
\Gluon(100,102)(100,18){2}{8}
\Text(15,30)[c]{$p_j-k_1-k_2$}
\Text(75,15)[c]{$p_j-k_1$}
\Text(110,5)[c]{$p_j$}
\LongArrow(110,65)(110,55)
\Text(120,60)[c]{$k_1$}
\end{picture}
\end{center}
\caption{\label{twoloop}  Two-loop diagram}
\end{figure}
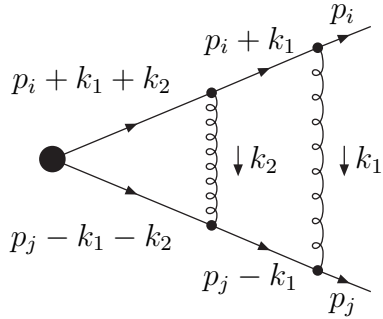

At two loops, the expression for $\omega_{ij}$ is 
\beqa   
\omega_{ij}^{(2)}(v_{i},v_{j})&=&
{g}_{s}^4 \int\frac{d^n k_1}{(2\pi)^n} \, \frac{(-i)}{k_1^2+i\epsilon} \,
\int\frac{d^n k_2}{(2\pi)^n} \, \frac{(-i)}{k_2^2+i\epsilon} \,
N^{\mu \nu}(k_1) \, N^{\rho \sigma}(k_2)
\nonumber \\ && \hspace{-30mm} \times \,
\frac{\Delta_{1i} \: v_{i}^{\mu}}
{\delta_{1i} v_{i} \cdot k_1+i\epsilon} \;
\frac{\Delta_{2i} \: v_{i}^{\rho}}
{\delta_{2i} v_{i} \cdot (k_1+k_2)+i\epsilon} \;
\frac{\Delta_{2j} \:v_{j}^{\sigma}} 
{\delta_{2j}v_{j} \cdot (k_1+k_2)+i\epsilon} \; 
\frac{\Delta_{1j} \:v_{j}^{\nu}} 
{\delta_{1j}v_{j} \cdot k_1+i\epsilon} \, .
\label{omega2}
\eeqa

We may rewrite (\ref{omega2}) as
\beqa  
\omega^{(2)}_{ij}(v_{i},v_{j})&=& 
{\cal S}_{ij}^{(2)}
\left[I_{11}^{(2)}(v_i, v_j)
+I_{12}^{(2)}(v_i, v_j, n)+I_{21}^{(2)}(v_i, v_j, n)\right.
\nonumber \\ &&
{}+I_{22}^{(2)}(v_i,v_j, n)+I_{13}^{(2)}(v_i, v_j, n^2)
+I_{31}^{(2)}(v_i, v_j, n^2)
\nonumber \\ && 
\left. {}+I_{23}^{(2)}(v_i, v_j, n, n^2)+I_{32}^{(2)}(v_i, v_j, n, n^2)
+I_{33}^{(2)}(v_i, v_j, n^2)\right] \, ,
\label{omegaII}
\end{eqnarray}
where ${\cal S}_{ij}^{(2)}$ is an overall sign
\beq
{\cal S}_{ij}^{(2)}=\Delta_{1i} \: \Delta_{1j} \: \Delta_{2i} \: 
\Delta_{2j} \: \delta_{1i} \: \delta_{1j} \:
\delta_{2i} \: \delta_{2j} \, .
\eeq

We will discuss explicitly the integral $I_{11}^{(2)}$.
For the production of a heavy quark pair $Q \bar Q$, 
$v_i=v_Q$, $v_j=v_{\bar Q}$, we find 
\beqa
\hspace{-20mm} 
I_{11}^{(2)} &\equiv& g_s^4 \int \frac{d^n k_1}{(2\pi)^n} \, 
\frac{(-i)}{k_1^2+i\epsilon} \frac{v_i \cdot v_j}
{(v_i \cdot k_1+i\epsilon) (v_j \cdot k_1+i\epsilon)}
\int \frac{d^n k_2}{(2\pi)^n} \, 
\frac{(-i)}{k_2^2+i\epsilon} \frac{v_i \cdot v_j}
{[v_i \cdot (k_1+k_2)+i\epsilon] [v_j \cdot (k_1+k_2)+i\epsilon]}
\nonumber \\ &=&
g_s^4 \int \frac{d^n k_1}{(2\pi)^n} \, 
\frac{i\, \pi^{\varepsilon/2} 2^{2\varepsilon} \, \Gamma(1+\varepsilon/2)\,
(1-2m^2/s)^2}{4 \pi^2 \; (k_1^2+i\epsilon) \; 
(v_i \cdot k_1+i\epsilon) \; (v_j \cdot k_1+i\epsilon)}
\left[\frac{1}{\varepsilon} F(x)
+\int_0^1 dx \int_0^1 dz \frac{f(x,z)}{(1-z)_+} \right]
\eeqa
where we used $v_i \cdot v_j=1-2m^2/s$, with $m$
the mass of the heavy quark. Here
\beqa
f(x,z)&=&\left[\frac{2m^2}{s}x^2+\frac{2m^2}{s}z^2(1-x)^2
+2\left(1-\frac{2m^2}{s}\right)x(1-x)z \right.
\nonumber \\ && \quad \left.
-4(1-x)(1-z) \, k_1\cdot \left(xv_i+z(1-x)v_j\right)\right]^{-1}
\eeqa
and 
\beq
F(x)=\int_0^1 dx \, f(x,1)=- \frac{L_{\beta}}{1-2m^2/s} 
\eeq
with $L_{\beta}$ defined in Eq. (\ref{Lb}).

After some further calculation we can extract the leading (double)
UV pole in the above integral in a simple form:
\beqa
I_{11}^{(2), \, {\rm UV \, DP}}&=&\frac{\alpha_s^2}{\pi^2}
\frac{1}{\varepsilon^2} L_{\beta}^2 \, .
\eeqa

Note that infrared poles are also encountered in this integral.
We can extract, for example, a term involving an ultraviolet and
an infrared pole:
\beq  
\frac{\alpha_s^2}{\pi^2}\frac{1}{\varepsilon} \frac{1}{\varepsilon^{\rm IR}}
L_{\beta} \frac{\left(1-2m^2/s\right)}{2\beta}
\ln\left(\frac{4m^2-2s-2s\beta}{4m^2-2s+2s\beta}\right)
=-\frac{\alpha_s^2}{\pi^2}\frac{1}{\varepsilon} \frac{1}{\varepsilon^{\rm IR}}
 L_{\beta} \, {\rm Re}\,L_{\beta} \,.
\eeq
However, here we are not interested in infrared poles or finite terms
so we will only concern ourselves with the UV poles at two loops.
As we see, the double UV poles take a particularly simple form.

After calculating the other $I^{(2)}$ integrals,
we can present the leading UV poles for the two-loop diagram in Figure 3:
\beqa
\omega_{ij}^{(2)\; \rm{UV \, DP}}(v_{i},v_{j})&=&
{\cal S}_{ij}^{(2)} \, \frac{\alpha_{s}^2}{\pi^2}
\frac{1}{\varepsilon^2}
\left[L_{\beta}^2 + 2L_{\beta}(L_i + L_j) -2L_{\beta}
+(L_i + L_j)^2-2(L_i + L_j) +1 \right]
\nonumber \\
&=&{\cal S}_{ij}^{(2)} \, \frac{\alpha_{s}^2}{\pi^2}
\frac{1}{\varepsilon^2} \left[L_{\beta}+L_i + L_j-1\right]^2 \,.
\label{omega2heavy}
\eeqa

More details and further results will appear in a forthcoming paper.

\end{document}